\documentclass[aps,prl,reprint,preprintnumbers]{revtex4-1}

\usepackage{amsmath}
\usepackage{amsfonts}
\usepackage{amssymb}
\usepackage{graphicx}
\usepackage{natbib}

\usepackage{color}
\usepackage[table]{xcolor}

\def\e{\epsilon}
\def\pol{\varepsilon}
\begin{document}

\preprint{SLAC--PUB--17410}

\title{\bf Graviton Bending in Quantum Gravity from One-Loop Amplitudes}

\author{Huan-Hang Chi$^{1,2}$}

\affiliation{$^1$SLAC National Accelerator Laboratory, Stanford University, 
Stanford, California 94309, USA \\
$^2$Physics Department, Stanford University, Stanford, California 94309, USA}


\begin{abstract}
We study the bending of gravitons that pass near a massive object like the Sun, using scattering amplitudes in which the Sun is represented by a massive scalar particle.  Our results complete previous work on the bending angles of massless spin-0, spin-$\frac{1}{2}$ and spin-1 particles~\cite{PhotonScalarBendingPRL,fermionBending,MoreCuts}, and provides more evidence for the violation of the equivalence principle at the quantum level, in the sense that the quantum corrections to bending angles for massless particles with different spins are different.  We provide a universal expression for the bending angle in terms of coefficients of triangle and bubble integrals in the amplitudes in the low energy limit.  We also compare bending angles for scalar, photon and graviton projectiles under different circumstances.
\end{abstract}


\maketitle

\bigskip

{\it Introduction} \ 
General relativity and quantum mechanics have vastly different foundations. The former requires a smooth spacetime and locality is absolute, while the latter requires regulation of short spacetime distances, and locality is inherently smeared by the uncertainty principle.  Trying to combine these two theories yields ultraviolet divergences that apparently can only be absorbed by an infinite set of counter-terms.  This situation is often referred to as a loss of predictivity for quantum gravity.  In gravitational theories with matter, such issues begin at one loop~\cite{tHooft:1974toh}.  For pure Einstein gravity, the loss of predictivity begins at two loops, as demonstrated by the (non-zero) universal renormalization scale dependence revealed in refs.~\cite{Bern:2015xsa,Bern:2017puu,Dunbar:2017qxb}, after earlier work on the (unphysical) two-loop divergence in dimensional regularization~\cite{Goroff:1985sz,Goroff:1985th,vandeVen:1991gw}. Although the ultraviolet properties of a complete theory of quantum gravity are still very unclear, we can nevertheless try to extract its long-range behavior, or infrared properties, by treating the quantum field theory of gravity as an effective theory and focusing on the long-range behavior~\cite{Donoghue:1994dn,BjerrumBohr:2002ks,BjerrumBohr:2002kt,Khriplovich:2002bt,Burgess:2003jk,Holstein:2008sx,OnShellTech,Bjerrum-Bohr:2015vda,Bern:2019nnu,Cachazo:2017jef,Guevara:2017csg,Xu:2017luq}.

The bending angle of light, or of any massless projectile, when it passes near a massive object such as the Sun, is a good observable to study to see how an effective theory of quantum gravity really works.  The massive object can be represented by a massive scalar particle. Using modern amplitude techniques, the bending angles for massless scalar and photon projectiles were calculated in ref.~\cite{PhotonScalarBendingPRL} and then extended to the massless fermion case in ref.~\cite{fermionBending}.  There the discontinuities of one-loop amplitudes with only gravitons crossing the cut were calculated and translated into a one-loop correction to the bending angle. Later, using traditional Feynman rules, contributions from scalars (photons) crossing the cut were also computed for the case of a scalar (photon) projectile~\cite{MoreCuts}.  These papers found the expected classical post-Newtonian correction to the bending angle for any projectile, while the quantum corrections differ for particles with different spins. The latter property indicates a violation of the classical equivalence principle at the quantum level.

In this paper, we complete this line of research by computing the bending angle for a graviton projectile.  Again we find the expected classical post-Newtonian correction to the bending angle, but a different quantum correction from what was found in the scalar and photon cases.  Our work provides more evidence for the violation of the classical principle of equivalence at the quantum level.  Our result for the graviton bending angle has more in common with ref.~\cite{MoreCuts}, in which the scalar (photon) cut contribution is included for scalar (photon) bending, than with ref.~\cite{PhotonScalarBendingPRL} in which only gravitons crossing the cut contribute.

This paper is organized as follows. First we list all tree-level amplitudes relevant for our graviton bending calculation.  Such amplitudes, even with external massive scalars, can be derived with some care from the four-point maximally-helicity-violating (MHV) amplitude of $\mathcal{N}=8$ supergravity (SUGRA).  Next we use the method of (generalized) unitarity~\cite{Bern:1994zx,Bern:1994cg,Bern:1997sc,Britto:2004nc} to fuse tree amplitudes into a one-loop amplitude. We then perform a standard tensor-integral reduction in order to write the one-loop amplitude in terms of scalar integrals. After taking the low-energy or long-range limit, the one-loop amplitude can be translated into a semiclassical effective potential.  Finally, we extract the one-loop correction to the bending angle from the potential using a semiclassical formula for angular deflection~\cite{Donoghue:1986ya}.  In addition to the pure-graviton case, we also provide a universal expression for the bending angle for scalar, photon and graviton projectiles in different setups, in terms of the amplitudes' scalar integral coefficients in the low energy limit.  We compare the various values of the bending angle, and comment briefly on the possible origin of the violation of classical equivalence principle found here and also in refs.~\cite{PhotonScalarBendingPRL,fermionBending,MoreCuts}.

\bigskip     

{\it Constructing the Loop Integrand} \ 
We consider a process in which an incoming graviton with momentum $k_1$ scatters off a massive scalar target with momentum $k_4$, into an outgoing graviton ($-k_2$) and massive scalar ($-k_3$), where $k_1+k_2+k_3+k_4=0$ in our all-incoming conventions.
The only interactions are Einstein gravity minimally coupled to a scalar field,
\begin{eqnarray}\label{InitialAction}
S_{g}&=&\int\mathrm{d}^4x\sqrt{-g}\left[
\frac{-2}{\kappa^2}R
+ \frac{1}{2}g^{\mu\nu}\partial_{\mu}\Phi \partial_{\nu}\Phi
- \frac{1}{2}M^2\Phi^2\right] \,,~~~
\end{eqnarray}
where $\kappa^2=32 \pi G$ and the metric signature is $(+---)$.

\begin{figure}
	\includegraphics[width=\linewidth]{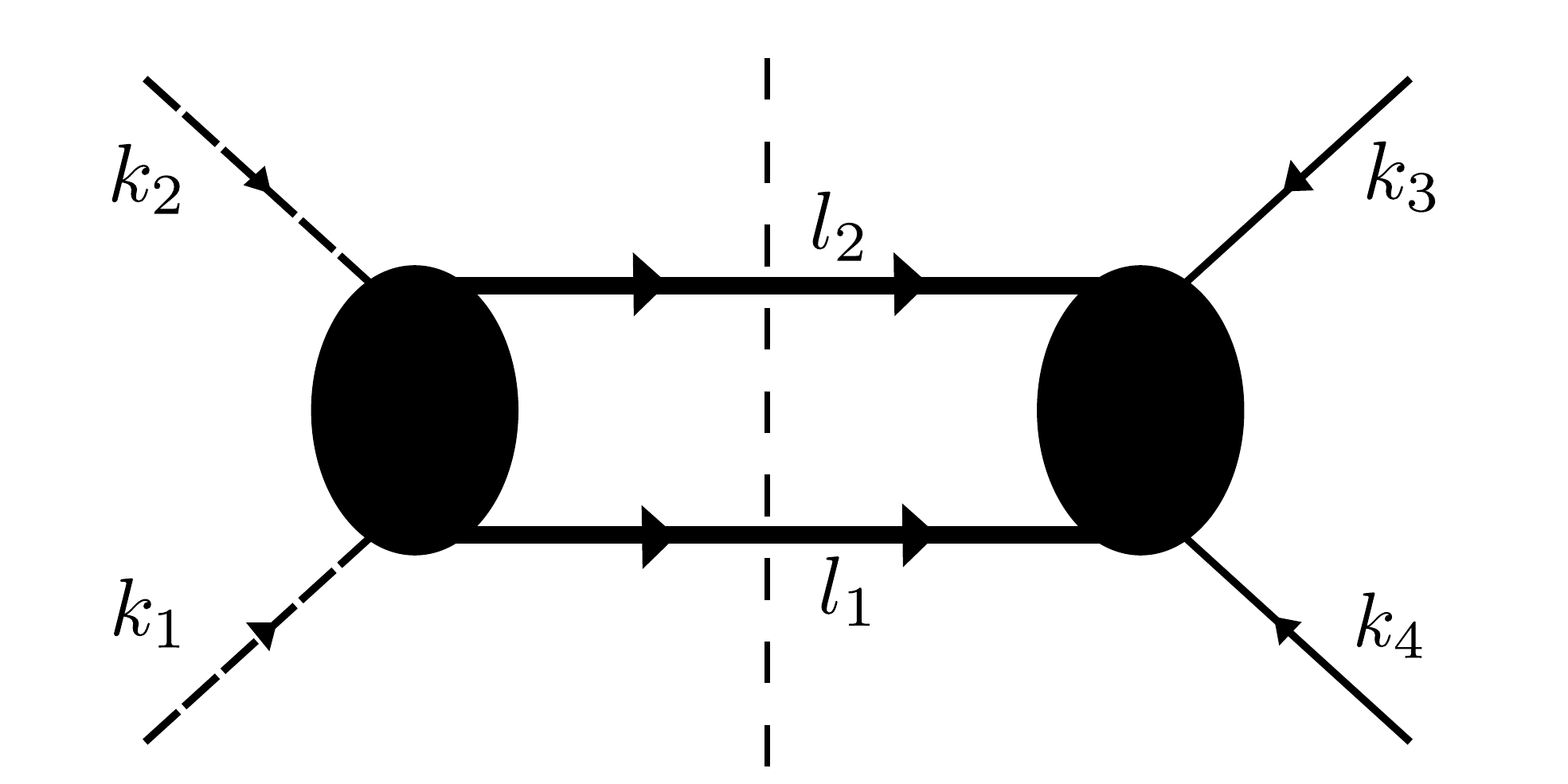}
	\caption{The unitarity cut for the scattering amplitude between the massless projectile (dashed line) and a massive scalar (solid line). We only have gravitons on the cut for the graviton bending, but we will include scalars (photons) crossing the cut for the case of scalar (photon) bending.}
	\label{fig:UniCut}
\end{figure}

We will build up the long-range part of the one-loop amplitude for this process, not from the action~(\ref{InitialAction}) but by considering cuts in the channel carrying the momentum-invariant $s \equiv (k_1+k_2)^2$. (This cut is called the $t$-channel cut in refs.~\cite{PhotonScalarBendingPRL,fermionBending,MoreCuts}.)  On the left side of the cut shown in Fig.~\ref{fig:UniCut} is a four-graviton tree amplitude; on the right side is a two-graviton two-massive-scalar tree amplitude.

Projectile helicity is conserved in forward scattering.
Taking all particles incoming, helicity selection rules imply that we need a four-graviton tree amplitude with two positive- and two negative-helicity gravitons, and a two-graviton-two-massive-scalar amplitude with opposite graviton helicities. These amplitudes are given by,  
\begin{eqnarray}\label{GravTree}
M_{[h^{+}(k_1) h^{-}(k_2)]}^{[h^{-}(l_2)h^{+}(l_1)]} &=& \frac{\kappa^2}{4} \frac{(2 k_1\cdot k_2)(2 k_2\cdot l_2) }{(2 k_1\cdot l_2)} \left(\frac{\langle 2 l_2\rangle^4}{PT}\right)^2  \,, \nonumber \\
M_{[h^{+}(k_1) h^{-}(k_2)]}^{[h^{+}(l_2)h^{-}(l_1)]} &=& \frac{\kappa^2}{4} \frac{(2 k_1\cdot k_2)(2 k_2\cdot l_2) }{(2 k_1\cdot l_2)} \left(\frac{\langle 2 l_1\rangle^4}{PT}\right)^2 \,, \nonumber\\
M_{[\Phi(k_4) \Phi(k_3)]}^{[h^{+}(l_2)h^{-}(l_1)]} &=& \frac{\kappa^2}{4} \frac{\langle l_1 | 3 | l_2]^2 \langle l_1 |4 | l_2]^2}{(2 l_1\cdot l_2)(2 l_1\cdot k_3)(2 l_1\cdot k_4)} \,, \nonumber \\
M_{[\Phi(k_4) \Phi(k_3)]}^{[h^{-}(l_2)h^{+}(l_1)]} &=& \frac{\kappa^2}{4} \frac{\langle l_2 | 3 | l_1]^2 \langle l_2 | 4 | l_1]^2}{(2 l_1\cdot l_2)(2 l_1\cdot k_3)(2 l_1\cdot k_4)} \,,
\end{eqnarray}
where $PT=\langle 1 2\rangle \langle 2 l_2 \rangle \langle l_2 l_1 \rangle \langle l_1 1\rangle$ is the Parke-Taylor factor. 
For convenience, we let graviton $1$\,(incoming) have positive helicity and graviton $2$\,(incoming) have negative helicity. To flip the helicities, we only need to take the complex conjugate of Eq.~(\ref{GravTree}) and other equations below.

Because the graviton helicity flips when crossing the cut, the one-loop amplitude reconstructed from its $s$-channel discontinuity is
\begin{eqnarray}\label{cut}
&&M_{[h^{+}(k_1)h^{-}(k_2)]}^{[\Phi(k_3) \Phi(k_4)]}\Big|_{1L} \nonumber \\
&=&\int\frac{d^D l}{(2\pi)^D}\frac{M_{[h^{+}(k_1) h^{-}(k_2)]}^{[h^{\mp}(-l_2)h^{\pm}(-l_1)]}M_{[\Phi(k_4) \Phi(k_3)]}^{[h^{\pm}(l_2)h^{\mp}(l_1)]}}{2 l_1^2 l_2^2} \nonumber \\
&=&\frac{\kappa^4}{32 s^4} \sum_{i=1}^{2} \sum_{j=3}^{4} \int\frac{d^D l}{(2\pi)^D} \frac{N^{+-}+N^{-+}}{l_1^2 l_2^2 P_i P_j}\,,
\end{eqnarray}
where $P_1=(k_1-l_1)^2,\, P_2=(k_2-l_1)^2,\, P_3=(k_3+l_1)^2-M^2,\, P_4=(k_4+l_1)^2-M^2$ and the numerator terms are
\begin{equation}\label{GNum}
N^{+-}+N^{-+}= \frac{[\text{tr}_{-}(1 3 2 l_2 3 l_1)]^4}{\langle 1 | 3 |2]^4} + (l_1 \leftrightarrow l_2) \,,
\end{equation}
where $\text{tr}_{-}(...) = \text{tr}(\frac{1}{2}(1-\gamma_5)...)$. After taking the fourth power in Eq.~(\ref{GNum}), we don't need to keep track of the part of the numerator that is linear in Levi-Civita tensor. After loop-momentum integration, this contraction can only generate $\pol_{\mu_1\mu_2\mu_3\mu_4}k_1^{\mu_1}k_2^{\mu_2}k_3^{\mu_3}k_4^{\mu_4}$, which vanishes due to momentum conservation and antisymmetry.

\bigskip

{\it Tensor Reduction} \ After performing a standard reduction of the one-loop tensor integrals in $D=4-2\e$ dimensions and combining contributions from the different pieces of Eq.~(\ref{cut}), we can write the amplitude in terms of scalar tadpole integrals, a scalar bubble integral, a scalar massless triangle integral, a scalar massive triangle integral and two scalar box integrals. Focusing on terms with an $s$-channel cut, we can discard the tadpoles and other irrelevant terms. The amplitude now reads
\begin{eqnarray}\label{AmplitudeExact}
\kappa^{-4} M_{[h^{+}(k_1)h^{-}(k_2)]}^{[\Phi(k_3) \Phi(k_4)]}\Big|_{1L}&=& b_1 I_4(s,t) + b_2 I_4(s,u) \nonumber \\
&&+ t_1 I_3(s) + t_2 I_3(s, M) \nonumber \\
&&+ b I_2(s) \,,
\end{eqnarray}
where $b_1$, $b_2$, $t_1$, $t_2$ and $b$ are polynomials in $s=(k_1+k_2)^2$ and $t=(k_2+k_3)^2$ as well as $M^2$.  For general $s$, $t$ and $M^2$, the exact forms of these coefficients are quite complicated. They can be found in the ancillary file {\tt Coeff-of-Integrals.txt}.  Also, $I_4(s,t)$ and $I_4(s,u)$ are scalar box integrals~\cite{OneLoopQCD} (where we made the $M$ dependence implicit for simplicity); $I_3(s)$ is the massless triangle integral; $I_3(s,M)$ is the finite massive triangle integral; and $I_2(s)$ is the massless scalar bubble integral~\cite{OnShellTech}.

\vspace*{3mm}

\bigskip

{\it Low Energy Limit} \ We take the energy of the external graviton $E=\omega$ to be much smaller than $M$, where $\omega$ is the frequency of the graviton, with $t=2M\omega + M^2$.  We take the transferred energy $s=-q^2$ to be even smaller, $s \ll \omega^2$.  In this low energy limit, the amplitude~(\ref{AmplitudeExact}) simplifies greatly.  To be precise, we first take the leading order terms in the $q^2$ expansion of the coefficients of the scalar integrals.  Higher order terms in $q^2$ will be a local term like $\delta(r)$, or derivatives of this $\delta$ function, after Fourier transformation to coordinate space.  Next we take the leading order terms in the expansion in $\omega$, and then set $D=4-2\e$ and take the $\epsilon^0$ and $\epsilon^1$ terms. We keep track of the $\epsilon^1$ terms since there might be some subtle $\frac{1}{\epsilon}\times\epsilon$ effects when combining the coefficients and the integrals that diverge as $\e\to0$. Now the one-loop amplitude reads
\begin{widetext}
	\begin{equation}\label{AmplitudeLE1}
	M_{[h^{+}(k_1)h^{-}(k_2)]}^{[\Phi(k_3) \Phi(k_4)]}|_{1L} = \frac{\kappa^4 N_g}{4} \Bigl[
	4(M \omega)^4(I_4(s,t)+ I_4(s,u)) + 4(M \omega)^2 s\, I_3(s) -\frac{15}{4}(M^2 \omega)^2 I_3(s, M)- \frac{29}{8}(M \omega)^2 I_2(s) \Bigr] \,,
	\end{equation}
\end{widetext}
where $N_g = (2M \omega)^4 / \langle 1 | 3 |2 ]^4$ is an overall phase factor.

The coefficients up to order $\epsilon^1$ are ($\frac{2947\epsilon}{360}-\frac{29}{8}, 4, \frac{\epsilon}{4}-\frac{15}{4}, 4$) respectively for bubble, massless triangle, massive triangle, and box.  Note that the coefficient of the massless triangle is $4$, without an order $\epsilon$ term. Thus when this coefficient is multiplied by the non-analytic divergent part of $I_3(s)$, $-\frac{\log(-s/\mu^2)}{s\epsilon}$, it will not give us contributions to the non-analytic pieces from the $\frac{1}{\epsilon}\times\epsilon$ effect.

The coefficient of the massive triangle is $-\frac{15}{4}$, which agrees with refs.~\cite{PhotonScalarBendingPRL, fermionBending}. The superficial difference of a factor of $\frac{1}{4}$ is due to a different definition of the massive triangle integral and won't affect the following results.

However, the coefficient of the massless triangle here is $4$, in contrast to $3$ in refs.~\cite{PhotonScalarBendingPRL, fermionBending}. This difference has a physical origin.  It comes from the fact that refs.~\cite{PhotonScalarBendingPRL, fermionBending} do not consider the contributions from scalars (photons) crossing the cut in the case of a scalar (photon) projectile.  These contributions are included in Figs.~2 and 3 in ref.~\cite{MoreCuts}. In that reference, the coefficient of the massless triangle is $4$, which agrees with our graviton bending result.  We will explore this point more later in this paper.

The coefficient of the bubble integral is $-\frac{29}{8}$, which differs from the corresponding coefficients in the scalar and photon cases. This result provides additional evidence that the classical equivalence principle is violated in some sense at the quantum level.

The scalar integrals appearing above can also be simplified in the low-energy limit. If we only keep terms with a cut in the $s$-channel and ignore the $1/\epsilon$ divergence~\cite{fermionBending}, we have
\begin{eqnarray}\label{scalarInt}
&& I_2(s) \simeq \frac{1}{16\pi^2}\left[-\log \left(-\frac{s}{\mu^2} \right)\right] \,, \nonumber\\
&& I_3(s) \simeq \frac{1}{16\pi^2}\left[-\frac{1}{2s}\log^2 \left(-\frac{s}{\mu^2} \right)\right] \,, \nonumber\\
&& I_3(s, M) \simeq \frac{1}{16\pi^2}\left[-\frac{1}{2M^2} \log \left(-\frac{s}{M^2} \right)-\frac{\pi^2}{2M\sqrt{-s}}\right] \,, \nonumber\\
&& I_4(s,t)+I_4(s,u) \simeq \frac{1}{16\pi^2}\left[\frac{2 i \pi }{M s \omega} \log \left(-\frac{s}{M^2} \right)\right] \,.
\end{eqnarray}
The scalar box part can be shown to exponentiate to an overall phase factor~\cite{Weinberg} and thus can be ignored. Then the amplitude can be rewritten as
%
%
\begin{widetext}
	\begin{equation}
	M_{[h^{+}(k_1)h^{-}(k_2)]}^{[\Phi(k_3) \Phi(k_4)]}\Big|_{1L}
	= \kappa^4 N_g (M \omega)^2 \biggl[
	\frac{15 \log \left(-\frac{s}{M^2}\right)}{512 \pi ^2}+\frac{15 M}{512
		\sqrt{-s}} 
	-\frac{\log ^2\left(-\frac{s}{\mu^2}\right)}{32 \pi ^2}+\frac{29 \log
		\left(-\frac{s}{\mu^2}\right)}{512 \pi ^2} \biggr] \,,
	\end{equation}
\end{widetext}
where again $s=-q^2$.

{\it From Low Energy Amplitude to Bending Angle} \  Next we extract the semiclassical potential via the Born approximation.  The following Fourier transformations will be useful
\begin{eqnarray}\label{FT}
&&\int \frac{d^3 q}{(2\pi)^3}e^{i \vec{q}\cdot \vec{r}} \frac{1}{q^2}=\frac{1}{4\pi r} \,, \nonumber \\
&&\int \frac{d^3 q}{(2\pi)^3}e^{i \vec{q}\cdot \vec{r}} \frac{1}{|q|}=\frac{1}{2\pi^2 r^2}  \,, \nonumber \\
&&\int \frac{d^3 q}{(2\pi)^3}e^{i \vec{q}\cdot \vec{r}} \log(q^2)=-\frac{1}{2\pi r^3} \,, \nonumber \\
&&\int \frac{d^3 q}{(2\pi)^3}e^{i \vec{q}\cdot \vec{r}} \log^2(\frac{q^2}{\mu^2})=\frac{2\log(\frac{r}{r_0})}{\pi r^3} \,,
\end{eqnarray}
where $r_0=e^{1-\gamma_E}\mu^{-1}$ and $\gamma_E$ is the Euler-Mascheroni constant. Here $r_0$ is related to infrared physics. It could be replaced by a more physical parameter by considering wavepackets for the external projectiles, instead of treating them as plane waves of infinite extent~\cite{Kosower:2018adc}, and taking into account finite detector resolution. But we won't do so here.

Now the one-loop semiclassical potential can be obtained as
\begin{eqnarray}
&V_{g}^{1L}(r)&\,\,=\frac{-1}{4M \omega}\int M_{[h^{+}(k_1)h^{-}(k_2)]}^{[\Phi(k_3) \Phi(k_4)]}|_{1L}(\vec{q}) e^{i \vec{q}\cdot \vec{r} / \hbar} \frac{d^3 q}{(2\pi)^3} \nonumber \\
&=&\frac{15 G^2 M^2 \omega}{-4 r^2} + \frac{G^2 M \omega \hbar  \left(16 \log \left(\frac{r}{r_0}\right)+11\right)}{\pi  r^3} \,. 
\end{eqnarray}

Next we can use the semiclassical formula~\cite{Donoghue:1986ya} for angular deflection to get the one-loop bending angle
\begin{eqnarray}
\theta_g^{1L} 
&=& \frac{b}{\omega}\int_{-\infty}^{+\infty} \frac{V_g^{\prime}(b\sqrt{1+u^2})}{\sqrt{1+u^2}} d u \nonumber \\
&=& \frac{15}{4}\frac{ G^2 M^2 \pi}{ b^2} + \frac{ \left(64 \log \left(\frac{2 r_0}{b}\right)-76\right)}{\pi} \frac{G^2 M \hbar}{b^3}\,,
\end{eqnarray}
%
%
where $V_g^{\prime} (r) \equiv dV_g^{1L}(r)/dr$
and $b$ is the gauge-invariant impact parameter.

The classical post-Newtonian correction is correctly reproduced, while the quantum correction is different from those of the scalar and photon cases~\cite{PhotonScalarBendingPRL, fermionBending, MoreCuts}. The most non-universal part of the result originates from the coefficient of the scalar bubble integral.  Our graviton-bending result provides more evidence of the violation of the classical equivalence principle in the sense that particles with different spins get different bending angles at the quantum level.

Following a similar procedure, the tree-level bending angle can be calculated easily using the tree 2-graviton-2-massive-scalar amplitude in Eq.~(\ref{GravTree}). It agrees with the tree-level scalar/photon/fermion bending. For completeness, we list it here as
\begin{equation}
\theta^{\text{tree}}_{g/\phi/\gamma/f} = \frac{4GM}{b}\,.
\end{equation}

Based on the above calculations, we can write down a general expression for the one-loop correction to the semiclassical potential and the bending angle:
\begin{widetext}
	\begin{equation}
	V^{1L}(r)=\frac{c_3 G^2 M^2 \omega}{r^2}+\frac{G^2 M \omega \hbar  \left(4 c_2 \log \left(\frac{r}{r_0}\right) - 2 c_1 - c_3\right)}{\pi  r^3}  \,,
	\end{equation}
	\begin{equation}\label{GeneralBA}
	\theta^{1L}=-\frac{  c_3 G^2 M^2 \pi}{b^2}+\frac{4 G^2 M \hbar  \left(4 c_2 \log (\frac{2 r_0}{b})+2 c_1-2 c_2+c_3\right)}{\pi  b^3}  \,,
	\end{equation} 
\end{widetext}
where again $b$ is the gauge-invariant impact parameter and $c_1$, $c_2$, $c_3$ are, respectively, the coefficients of the scalar massless bubble integral $I_2(s)$, the scalar massless triangle integral $I_3(s)$ and the scalar massive triangle integral $I_3(s, M)$ in Eq.~(\ref{AmplitudeLE1}).

\bigskip

{\it Comparison with Previous Results} \ 
Following a similar procedure, we have redone the previously-computed bending of scalar and photon projectiles.  Here we assume no scalar self-interactions, and every interaction includes gravitation. The action reads
\begin{widetext}
\begin{eqnarray}\label{Action}
S_{\phi}&=&\int\mathrm{d}^4x\sqrt{-g}\left[\frac{-2}{\kappa^2}R+\frac{1}{2} g^{\mu\nu} \partial_{\mu}\phi \partial_{\nu}\phi+\frac{1}{2}g^{\mu\nu}\partial_{\mu}\Phi \partial_{\nu}\Phi-\frac{1}{2}M^2\Phi^2\right]\,,\nonumber \\
S_{\gamma}&=&\int\mathrm{d}^4x\sqrt{-g}\left[\frac{-2}{\kappa^2}R-\frac{1}{4} g^{\mu\rho}g^{\nu\sigma} F_{\mu\nu}F_{\rho\sigma}+\frac{1}{2}g^{\mu\nu}\partial_{\mu}\Phi \partial_{\nu}\Phi-\frac{1}{2}M^2\Phi^2\right] \,.
\end{eqnarray}
\end{widetext}

There are two options for the calculation in each case:
\begin{enumerate}
\item only gravitons cross the cut in Fig.~\ref{fig:UniCut}, as in ref.~\cite{PhotonScalarBendingPRL};
\item both gravitons and scalars (photons) cross the cut in Fig.~\ref{fig:UniCut} for scalar (photon)
bending, as in ref.~\cite{MoreCuts}.
\end{enumerate}
The coefficients $c_1$, $c_2$, $c_3$ mentioned above for each case are re-calculated here using amplitude techniques, and are shown in Table~\ref{table:Coeff}. They agree with previous results on scalar and photon bending in refs.~\cite{PhotonScalarBendingPRL, fermionBending, MoreCuts}.

There are several subtleties worth mentioning about the tree amplitudes used here.  Tree amplitudes involving external gravitons can be obtained directly from e.g.~ref.~\cite{Berends:1988zp} (for pure gravity) or from the four-point MHV amplitude of $\mathcal{N}=8$ SUGRA (see the excellent review~\cite{Elvang:2013cua}) and that makes our life quite easy.
The self-interactions of matter fields in N=8 SUGRA don't contribute to the 4-point tree amplitudes with external gravitons used here.

And indeed for case 1, i.e., only only gravitons cross the cut in Fig.~\ref{fig:UniCut}, using supersymmetry(SUSY),the numerator in Eq.(\ref{cut}) for a spin-$j$ projectile can be shown to be
\begin{eqnarray}\label{generalNum}
&&N^{(j)} = \left(\frac{\langle 1 l_2 \rangle}{\langle 2 l_2 \rangle}\right)^{2(2-j)} N^{+-} + (l_1 \leftrightarrow l_2) \nonumber \\
&=& \left( \langle 1 |3|2] \frac{\text{tr}_{-}(1 l_2 3 l_1)}{\text{tr}_{-}(1 3 2 l_2 3 l_1)}\right)^{2(2-j)} N^{+-} + (l_1 \leftrightarrow l_2) \nonumber \\
&=& \frac{[\text{tr}_{-}(1 l_2 3 l_1)]^{4-2j} [\text{tr}_{-}(1 3 2 l_2 3 l_1)]^{2j}}{\langle 1 |3|2]^{2j}} + (l_1 \leftrightarrow l_2) ,
\end{eqnarray}
where $N^{+-}$ is the (partial) numerator of graviton bending in Eq.(\ref{cut},\,\ref{GNum}). Taking $j=0,\frac{1}{2}, 1$, the numerator given in Eq.(\ref{generalNum}) agree with those in \cite{PhotonScalarBendingPRL,fermionBending}. Now the calculation for different projectiles with only graviton crossing the cut is straightforward.

However, to calculate the scalar (photon) loop contribution for the scalar (photon) bending, we need tree amplitudes without an external graviton.  There are such amplitudes in $\mathcal{N}=8$ SUGRA, however, we can't use them because they could contain contributions, either from undesired mediating particles from the $\mathcal{N}=8$ SUGRA multiplet, or else undesired contact terms, such as the sigma model term for the four-scalar tree amplitude.  Also, we can't use BCFW recursion relations for these amplitudes, which is again related to possible undesired contact terms.  Hence, to get these four-point tree amplitudes without external gravitons, we use the three-point Feynman rules listed in Eqs.~(4.10)-(4.12) in ref.~\cite{Bjerrum-Bohr:2014lea}. By connecting them with a graviton propagator (or a graviton projector), we can get the desired tree amplitudes with a mediating graviton as
\begin{eqnarray}\label{FeynTree}
&&M_{[\phi(k_1) \phi(k_2)]}^{[\phi(k_3)\phi(k_4)]} = \frac{\kappa^2}{4} \frac{(s^2+s t+t^2)^2}{stu} \,,\nonumber\\
&&M_{[\phi(k_1) \phi(k_2)]}^{[\Phi(k_3)\Phi(k_4)]} = \frac{\kappa^2}{4} \frac{(t-M^2)(u-M^2)}{s} \,,\nonumber\\
&&M_{[\gamma^{+}(k_1) \gamma^{-}(k_2)]}^{[\gamma^{+}(k_3)\gamma^{-}(k_4)]} = -\frac{\kappa^2}{4} \langle 24 \rangle^2 [13]^2 (\frac{1}{s}+\frac{1}{t})\,,\nonumber\\
&&M_{[\gamma^{+}(k_1) \gamma^{-}(k_2)]}^{[\gamma^{-}(k_3)\gamma^{+}(k_4)]} = -\frac{\kappa^2}{4} \langle 23 \rangle^2 [14]^2 (\frac{1}{s}+\frac{1}{u})\,,\nonumber\\
&&M_{[\gamma^{+}(k_1) \gamma^{-}(k_2)]}^{[\Phi(k_3)\Phi(k_4)]} = -\frac{\kappa^2}{4} \langle 2 |3|1]^2 \frac{1}{s}\,,\nonumber\\
&&M_{[\gamma^{-}(k_1) \gamma^{+}(k_2)]}^{[\Phi(k_3)\Phi(k_4)]} = -\frac{\kappa^2}{4} \langle 1 |3|2]^2 \frac{1}{s}\,,
\end{eqnarray}
where again $s=(k_1+k_2)^2$, $t=(k_1+k_4)^2$, $u=(k_1+k_3)^2$ and all momenta are incoming. Our expression for the four-massless-scalar amplitude with a mediating graviton in Eq.~(\ref{FeynTree}) also agrees with Eq.~(2.54) in ref.~\cite{Pasukonis:2005db}.

\begin{table}
	\centering 
	\begin{tabular}{c c c c c} 
		\hline\hline 
		\,\,\,\, projectile \,\,\,\,&particles crossing cut  & \,\,\,\,$c_1$ \,\,\,\,& \,\,\,\,$c_2$ \,\,\,\,& \,\,\,\,$c_3$ \,\,\,\,\\ [0.5ex] 
		\hline 
		$\phi$ &$h$ & $\frac{3}{40}$ & 3 & $-\frac{15}{4}$ \\[1ex] 
		$\phi$ &$h$, $\phi$& $\frac{371}{120}$ & 4 & $-\frac{15}{4}$ \\[1ex]
		$\gamma$ &$h$ & $-\frac{161}{120}$ & 3 & $-\frac{15}{4}$\\[1ex]
		$\gamma$ &$h$, $\gamma$& $\frac{113}{120}$ & 4 & $-\frac{15}{4}$ \\[1ex]
		$h$&$h$& $-\frac{29}{8}$ & 4 & $-\frac{15}{4}$\\ [1ex] 
		\hline 
	\end{tabular}
	\caption{Projectile bending at the quantum level for a variety of theories.  $c_1$, $c_2$, $c_3$ are, respectively, the coefficients of the scalar massless bubble integral $I_2(s)$, the scalar massless triangle integral $I_3(s)$ and the scalar massive triangle integral $I_3(s, M)$ in Eq.~(\ref{AmplitudeLE1}). In each case, the first column denotes the projectiles, while the second column denotes the particles crossing the cut.} 
	\label{table:Coeff} 
\end{table}

Note that by including contributions from scalars (photons) crossing the cut in the scalar (photon) bending case, the value of $c_2$ rise from 3 to 4, in agreement with the graviton-bending case.  Thus, by including every possible gravitational interaction in each action in Eq.~(\ref{Action}), we obtain a greater level of agreement in the bending angle of particles with different spins.  More precisely, the product of the coefficient $c_2$ and the infrared-divergent massless triangle integral $I_3(s)$ gives us the corresponding infrared divergence. There are two types of infrared divergences:
\begin{enumerate}
\item a soft divergence when one of the two gravitons crossing the cut becomes soft;
\item a forward scattering pole when the angle between the projectile and the particle crossing the cut (when it is the same particle as the projectile) becomes small.
\end{enumerate}
Both divergences are universal.  The first generates the ``3'' for $c_2$, while the second generates an additional ``1'' in the table.
When scalars (photons) are included on the cut for the scalar (photon) bending case, the forward-scattering pole is properly taken into account.  Then the infrared divergence rises by one unit in $c_2$ and matches that of the graviton bending case, where the two types of infrared divergences arise together, with only having gravitons on the cut.

The remaining non-universal difference comes from the scalar bubble integral coefficient, $c_1$. This difference indicates the violation of the classical equivalence principle at the quantum level.

Given the values of $c_1$, $c_2$, $c_3$ in Table \ref{table:Coeff}, we can use Eq.~(\ref{GeneralBA}) to calculate the one-loop correction to the bending angle in each case.  For example, we substitute the second and fourth row of Table \ref{table:Coeff} into Eq.~(\ref{GeneralBA}) and get the scalar (photon) bending angle with both scalar (photon) and graviton crossing the cut,
\begin{equation}\label{spBAcomplete}
\theta^{1L}_{\phi/\gamma} = \frac{15  G^2 M^2 \pi}{4 b^2}
+ \frac{G^2 M \hbar  \left(64 \log \left(\frac{2 r_0}{b}\right)+8 c_1-47\right)}{\pi  b^3} \,,
\end{equation}  
where $c_1= \frac{371}{120}$, $\frac{113}{120}$ for scalar and photon, respectively. This result agrees perfectly with ref.~\cite{MoreCuts}, and also verifies the correctness of our general expression Eq.~(\ref{GeneralBA}) for one-loop correction to bending angles.

Similarly, when we substitute the first and third row of Table \ref{table:Coeff} into Eq.~(\ref{GeneralBA}), we get the scalar (photon) bending angle with only gravitons crossing the cut as
\begin{equation}\label{spBApartial}
\theta^{1L}_{\phi/\gamma} = \frac{15 G^2 M^2 \pi}{4 b^2}
+ \frac{G^2 M \hbar \left(48 \log \left(\frac{2 r_0}{b}\right)+8 c_1-39\right)}{\pi  b^3} \,,
\end{equation}  
where $c_1= \frac{3}{40}$, $-\frac{161}{120}$ for scalar and photon, respectively.  This expression differs from Eq.~(12) in ref.~\cite{PhotonScalarBendingPRL}. If we were to take $c_2 \rightarrow -c_2$ in Eq.~(\ref{GeneralBA}), this difference would go away. Since our general expression~(\ref{GeneralBA}) is verified by the results in ref.~\cite{MoreCuts}, it seems likely that ref.~\cite{PhotonScalarBendingPRL} has missed a minus sign in the Fourier transformation of $\log^2(\frac{q^2}{\mu^2})$ in Eq.~(\ref{FT}). This sign has propagated to ref.~\cite{fermionBending}.

\bigskip

{\it Equivalence Principle violated?} \  As shown in refs.~\cite{PhotonScalarBendingPRL, fermionBending} for spin-0, spin-$\frac{1}{2}$ and spin-1 particles, and now here for 
spin-2 gravitons, massless particles don't follow the same geodesic anymore at the one-loop level and have different bending angles.  These results show that the classical equivalence principle is violated in some sense. However, at the quantum level, massless particles have a wavelength and cannot really be treated as point particles. So there should be a tidal force on them, given that the gravitational field is not uniform.  Thus the violation of the equivalence principle we see here might just be an effect of this non-locality, which imposes no real challenge to the equivalence principle. If we let the wavelength of the massless particles be much shorter (but not too short so we don't need to care about the ultraviolet details of quantum gravity) than the radius of curvature, the difference in bending angles might go away. We are not very sure about this and will hopefully explore it in further work.


{\it Acknowledgments} \ We thank Lance Dixon for inspiring and useful discussions and a careful reading of the draft, Leonard Susskind for comments on the violation of the equivalence principle, and N. E. J. Bjerrum-Bohr and Pierre Vanhove for a careful reading of the draft and useful comments. This work is supported by the U.S. Department of Energy (DOE) under contract DE-AC02-76SF00515.


\begin{thebibliography}{99}

\bibitem{PhotonScalarBendingPRL} 
N.~E.~J.~Bjerrum-Bohr, J.~F.~Donoghue, B.~R.~Holstein, L.~Planté and P.~Vanhove,
Phys.\ Rev.\ Lett.\  {\bf 114}, no. 6, 061301 (2015)
doi:10.1103/PhysRevLett.114.061301
[arXiv:1410.7590 [hep-th]].

\bibitem{fermionBending} 
N.~E.~J.~Bjerrum-Bohr, J.~F.~Donoghue, B.~R.~Holstein, L.~Plante and P.~Vanhove,
JHEP {\bf 1611}, 117 (2016)
doi:10.1007/JHEP11(2016)117
[arXiv:1609.07477 [hep-th]].

\bibitem{MoreCuts} 
D.~Bai and Y.~Huang,
Phys.\ Rev.\ D {\bf 95}, no. 6, 064045 (2017)
doi:10.1103/PhysRevD.95.064045
[arXiv:1612.07629 [hep-th]].

\bibitem{tHooft:1974toh} 
G.~'t Hooft and M.~J.~G.~Veltman,
Ann.\ Inst.\ H.\ Poincare Phys.\ Theor.\ A {\bf 20}, 69 (1974).

\bibitem{Bern:2015xsa} 
Z.~Bern, C.~Cheung, H.~H.~Chi, S.~Davies, L.~Dixon and J.~Nohle,
Phys.\ Rev.\ Lett.\  {\bf 115}, no. 21, 211301 (2015)
doi:10.1103/PhysRevLett.115.211301
[arXiv:1507.06118 [hep-th]].

\bibitem{Bern:2017puu} 
Z.~Bern, H.~H.~Chi, L.~Dixon and A.~Edison,
Phys.\ Rev.\ D {\bf 95}, no. 4, 046013 (2017)
doi:10.1103/PhysRevD.95.046013
[arXiv:1701.02422 [hep-th]].

\bibitem{Dunbar:2017qxb} 
D.~C.~Dunbar, G.~R.~Jehu and W.~B.~Perkins,
Phys.\ Rev.\ D {\bf 95}, no. 4, 046012 (2017)
doi:10.1103/PhysRevD.95.046012
[arXiv:1701.02934 [hep-th]].

\bibitem{Goroff:1985sz} 
M.~H.~Goroff and A.~Sagnotti,
Phys.\ Lett.\  {\bf 160B}, 81 (1985).
doi:10.1016/0370-2693(85)91470-4

\bibitem{Goroff:1985th} 
M.~H.~Goroff and A.~Sagnotti,
Nucl.\ Phys.\ B {\bf 266}, 709 (1986).
doi:10.1016/0550-3213(86)90193-8

\bibitem{vandeVen:1991gw} 
A.~E.~M.~van de Ven,
Nucl.\ Phys.\ B {\bf 378}, 309 (1992).
doi:10.1016/0550-3213(92)90011-Y

\bibitem{Donoghue:1994dn} 
J.~F.~Donoghue,
Phys.\ Rev.\ D {\bf 50}, 3874 (1994)
doi:10.1103/PhysRevD.50.3874
[gr-qc/9405057].

\bibitem{BjerrumBohr:2002ks} 
N.~E.~J.~Bjerrum-Bohr, J.~F.~Donoghue and B.~R.~Holstein,
Phys.\ Rev.\ D {\bf 68}, 084005 (2003)
Erratum: [Phys.\ Rev.\ D {\bf 71}, 069904 (2005)]
doi:10.1103/PhysRevD.68.084005, 10.1103/PhysRevD.71.069904
[hep-th/0211071].

\bibitem{BjerrumBohr:2002kt} 
N.~E.~J.~Bjerrum-Bohr, J.~F.~Donoghue and B.~R.~Holstein,
Phys.\ Rev.\ D {\bf 67}, 084033 (2003)
Erratum: [Phys.\ Rev.\ D {\bf 71}, 069903 (2005)]
doi:10.1103/PhysRevD.71.069903, 10.1103/PhysRevD.67.084033
[hep-th/0211072].


\bibitem{Khriplovich:2002bt} 
I.~B.~Khriplovich and G.~G.~Kirilin,
J.\ Exp.\ Theor.\ Phys.\  {\bf 95}, no. 6, 981 (2002)
[Zh.\ Eksp.\ Teor.\ Fiz.\  {\bf 122}, no. 6, 1139 (2002)]
doi:10.1134/1.1537290
[gr-qc/0207118].

\bibitem{Burgess:2003jk} 
C.~P.~Burgess,
Living Rev.\ Rel.\  {\bf 7}, 5 (2004)
doi:10.12942/lrr-2004-5
[gr-qc/0311082].

\bibitem{Holstein:2008sx} 
B.~R.~Holstein and A.~Ross,
arXiv:0802.0716 [hep-ph].

\bibitem{OnShellTech} 
N.~E.~J.~Bjerrum-Bohr, J.~F.~Donoghue and P.~Vanhove,
JHEP {\bf 1402}, 111 (2014)
doi:10.1007/JHEP02(2014)111
[arXiv:1309.0804 [hep-th]].

\bibitem{Bjerrum-Bohr:2015vda} 
N.~E.~J.~Bjerrum-Bohr, J.~F.~Donoghue, B.~K.~El-Menoufi, B.~R.~Holstein, L.~Planté and P.~Vanhove,
Int.\ J.\ Mod.\ Phys.\ D {\bf 24}, no. 12, 1544013 (2015)
doi:10.1142/S0218271815440137
[arXiv:1505.04974 [hep-th]].

\bibitem{Bern:2019nnu} 
Z.~Bern, C.~Cheung, R.~Roiban, C.~H.~Shen, M.~P.~Solon and M.~Zeng,
arXiv:1901.04424 [hep-th].

\bibitem{Cachazo:2017jef} 
F.~Cachazo and A.~Guevara,
arXiv:1705.10262 [hep-th].

\bibitem{Guevara:2017csg} 
A.~Guevara,
arXiv:1706.02314 [hep-th].

\bibitem{Xu:2017luq} 
C.~Xu and Y.~Yang,
J.\ Math.\ Phys.\  {\bf 59}, no. 3, 032501 (2018)
doi:10.1063/1.5009911
[arXiv:1709.04127 [gr-qc]].

\bibitem{Bern:1994zx} 
Z.~Bern, L.~J.~Dixon, D.~C.~Dunbar and D.~A.~Kosower,
Nucl.\ Phys.\ B {\bf 425}, 217 (1994)
doi:10.1016/0550-3213(94)90179-1
[hep-ph/9403226].

\bibitem{Bern:1994cg} 
Z.~Bern, L.~J.~Dixon, D.~C.~Dunbar and D.~A.~Kosower,
Nucl.\ Phys.\ B {\bf 435}, 59 (1995)
doi:10.1016/0550-3213(94)00488-Z
[hep-ph/9409265].

\bibitem{Bern:1997sc} 
Z.~Bern, L.~J.~Dixon and D.~A.~Kosower,
Nucl.\ Phys.\ B {\bf 513}, 3 (1998)
doi:10.1016/S0550-3213(97)00703-7
[hep-ph/9708239].

\bibitem{Britto:2004nc} 
R.~Britto, F.~Cachazo and B.~Feng,
Nucl.\ Phys.\ B {\bf 725}, 275 (2005)
doi:10.1016/j.nuclphysb.2005.07.014
[hep-th/0412103].

\bibitem{Donoghue:1986ya} 
J.~F.~Donoghue and B.~R.~Holstein,
Am.\ J.\ Phys.\  {\bf 54}, 827 (1986).
doi:10.1119/1.14423

\bibitem{OneLoopQCD} 
R.~K.~Ellis and G.~Zanderighi,
JHEP {\bf 0802}, 002 (2008)
doi:10.1088/1126-6708/2008/02/002
[arXiv:0712.1851 [hep-ph]].

\bibitem{Weinberg} 
S.~Weinberg,
Phys.\ Rev.\  {\bf 140}, B516 (1965).
doi:10.1103/PhysRev.140.B516

\bibitem{Kosower:2018adc} 
D.~A.~Kosower, B.~Maybee and D.~O'Connell,
JHEP {\bf 1902}, 137 (2019)
doi:10.1007/JHEP02(2019)137
[arXiv:1811.10950 [hep-th]].

\bibitem{Berends:1988zp} 
F.~A.~Berends, W.~T.~Giele and H.~Kuijf,
Phys.\ Lett.\ B {\bf 211}, 91 (1988).
doi:10.1016/0370-2693(88)90813-1

\bibitem{Elvang:2013cua} 
H.~Elvang and Y.~t.~Huang,
arXiv:1308.1697 [hep-th].

\bibitem{Bjerrum-Bohr:2014lea} 
N.~E.~J.~Bjerrum-Bohr, B.~R.~Holstein, L.~Planté and P.~Vanhove,
Phys.\ Rev.\ D {\bf 91}, no. 6, 064008 (2015)
doi:10.1103/PhysRevD.91.064008
[arXiv:1410.4148 [gr-qc]].

\bibitem{Pasukonis:2005db} 
J.~Pasukonis,
Fortsch.\ Phys.\  {\bf 53}, 1011 (2005)
doi:10.1002/prop.200510249
[hep-th/0506065].


\end{thebibliography}

\end{document}